\documentclass[11pt]{article}
\usepackage{graphicx}
\usepackage{amsmath}
\date{\today}
\begin{document}
\sloppy
\title{Two-component universe containing ordinary mater and a new component of the form $\frac{l}{R}\rho$}
\author{V. Majern\'{\i}k\\
Institute of Mathematics Slovak Academy of Sciences,\\ SK-814 73
Bratislava, \v Stef\'anikova 49, Slovak Republic}
\maketitle

\begin{abstract}
We consider a model of the universe described by Einstein equations whose the right-hand side consists of ordinary energy-momentum tensor and an effective vacuum energy-momentum tensor with $\rho^{vac}=-p^{vac}=L_c/K$, where $L_c=8\pi Gl\rho R^{-1}$ and $\rho$ is the mass density of cosmical medium,
$l$ the constat with dimension $[cm]$ and $R$ the scale factor of the expanding universe.
We determine the vacuum energy density assigned to the term $L_c$ as a function of the scale factor and, when taking
$l=2R_0$, $R_0$ being the present 'radius' of the universe, and $\Omega_0=0.3$, the normalized vacuum mass density, we obtain $\Omega_t=1$, deceleration parameter $q_0\approx -0.5$ and the age of the universe  $t_0\approx 0.9\times H_0^{-1}$.
The values are compatible with the estimated observatory values of cosmic parameters. Finally, we try to interpret the energy content of $L_c$
as the field energy of a field whose source is the ordinary matter in the universe.

\end{abstract}

\section{Introduction}

Since the discovery of the accelerating expansion of the universe \cite{TO}, dark energy
is often invoked to explain this phenomena.
 The basic set of  astronomical observations confirming the existence of dark energy
includes: observations from SNe Ia, CMB anisotropies, large scale structure, X-ray data
from galaxy clusters, age estimates of globular clusters and old high red-shift galaxies (OHRG's).
These observatory results seem to supply  the remaining piece of information confirming the inflationary flatness
prediction $(\Omega_t\approx 1)$.
The existence of an extra component filling the universe
has also indirectly been suggested by independent
studies based on fluctuations of the 3K relict radiation
\cite{4}, large scale structure \cite{5} age estimates of globular
clusters or old high redshift objects \cite{6}, as well as by the
X-ray data from galaxy clusters \cite{7}.
These observations strongly suggest that the universe is  flat
composed of $\rho \sim 1/3$ of matter (baryonic + dark) and $\rho_{\lambda}\sim 2/3$ of an exotic component.

 In the framework of quantum field theory
the presence of dark energy is due to the zero-point energy of
all particles and fields filling the universe which manifests
itself in several quantum phenomena like the Lamb
shift and Casimir effect \cite{10}.
However, the conflict between the estimating
observatory value of dark energy and that of quantum field prediction \cite{WEI} inspired many authors
to find alternative for the field-theory model of dark energy.
The well-known alternative is the Einstein cosmological constant $\Lambda$
which  is a time independent and spatially uniform
dark component. It may be classically interpreted as
a relativistic perfect fluid obeying the equation of
state $p = -\rho$.
Recently, many
models with variable cosmological constant have been proposed, in some cases it
depends on scale factor \cite{SF},\cite{17} or on the cosmic time \cite{T} or on both of them (for an overview see e.g. \cite{MM}). Other authors have suggested
that the cosmological constant can be written as a trace of an energy-momentum
tensor \cite{M}.

In this report, we
study  the cosmic evolution supposing that the energy-momentum tensor in Einstein's equations consists of ordinary energy-momentum tensor and an effective vacuum energy-momentum tensor with $\rho^{vac}=-p^{vac}=L_c/K$, where $L_c=8\pi Gl\rho R^{-1}$ and $\rho$ is the mass density of cosmical medium, one $L_c=8\pi l\rho R^{-1}$, where $\rho$ is the mass density of the cosmic medium
$R$ is the scale factor and $l$ is a free constant having the length dimension.

\section{Investigate of cosmic evolution described by the modified energy-momentum tensor}

 We start with the Einstein equations
$$R^{ik}-(1/2)R=-K T^{ik}+L_c g^{ik} \qquad K=8\pi G.$$
 $L_c$ can be phenomenologically viewed as a dynamical cosmological term. Using the Robertson-Walker metric for flat space $(k=0)$ and
$L_c=Kl\rho R^{-1}$,
we have
\begin{equation} \label{1}
(\frac{\dot{R}}{R})^2=\frac{K}{3}(\frac{l\rho}{R})+ (\frac{K}{3} )\rho=
(\frac{K\rho}{3}) (1+\frac{l}{R})
\end{equation}
and
\begin{equation}  \label{2}
\frac{2\ddot{R}}{R}+(\frac{\dot{R}}{R})^2=-K p+\frac{Kl\rho}{R}.
\end{equation}
The corresponding total energy-conservation law reads \cite{17}
\begin{equation}  \label{3}
\frac{d}{dR}(\rho R^3)+3pR^2=-\frac{1}{K}[ \frac{d}{dR}(L_c R^3)-3L_c R^2]
=-\frac{1}{K}[\frac{d}{dR}(K\rho l R^2)-3Kl\rho R
].
\end{equation}
One can view $-(L_c/K) g^{ik}$ as the effective vacuum energy-momentum tensor with $\rho^{vac}=-p^{vac}=L_c/K$. The vacuum energy density assigned to $L_c $ is
\begin{equation} \label{4}
\rho_c=
(\frac{l}{R})\rho.
\end{equation}

For the radiation-dominated epoch ($p=1/3$), equation(3) leads to the following differential equation
\begin{equation} \label{5}
\frac{d\rho_r(R)}{dR}(R^3+lR^2)+\rho_r(R)(3R^2+R^2-lR)=0.
\end{equation}
Its solution is
\begin{equation} \label{6}
\rho_r(R)=\frac{C_1R}{(l+R)^5},
\end{equation}
where $C_1$ is a positive integration constant.
For the mater-dominated epoch $(p=0)$, equation(5) gets the form
\begin{equation} \label{7}
\frac{d\rho_m(R)}{dR}(R^3+lR^2)+\rho_m(R)(3R^2-lR)=0,
\end{equation}
whose solution is
\begin{equation} \label{8}
\rho_m(R)=\frac{C_2R}{(l+R)^4},
\end{equation}
where $C_2$ being again a positive integration constant.
The solutions of equation(6) and equation(8), $\rho_r(R)$ and $\rho_m(R)$, as  functions of $R$ are shown in Fig. 1.

\begin{figure}[h]
\includegraphics[scale=0.9]{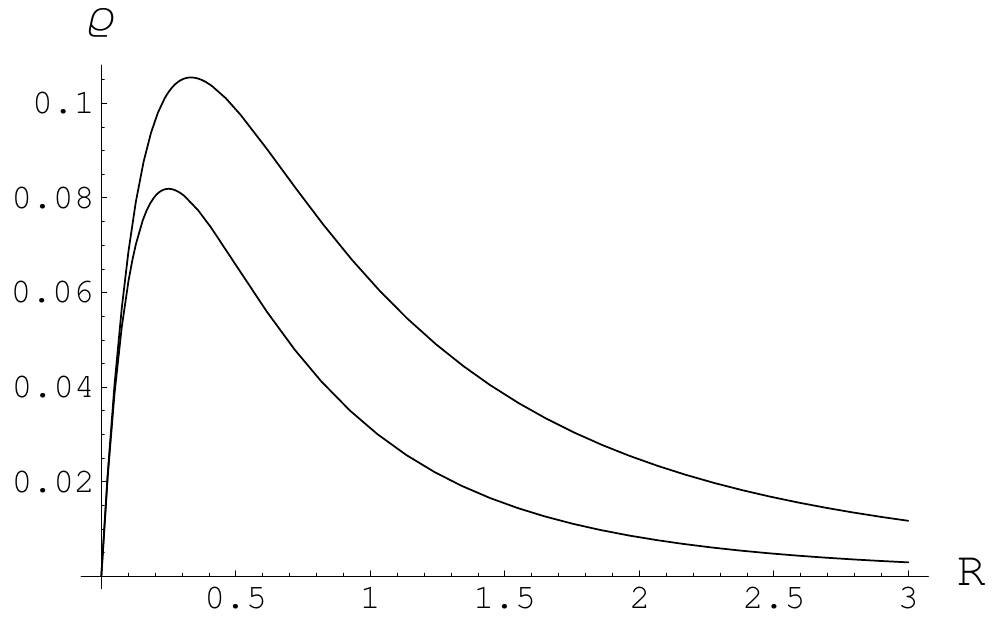}
\caption{$\rho $ as a function of $R$, $(l=1)$. The upper curve refers to $\rho_m$ and the lower to $\rho_r$.  }
\label{Rho1}
\end{figure}

One sees that their graphical representations
show one-hump curves whose maxima lie in $R=l/3$ for ($p$=0) and in $R=l/4$ for ($p$=1/3), respectively.

Knowing $\rho$ as a function of $R$ we can determine $R(t)$ for all time by solving equation(1) and equation(2). Substituting $\rho_r(R)$ or $\rho_m(R)$ into equations (\ref{1}) and (\ref{2}) we get for $p=1/3$ or $p=0$
\begin{equation} \label{9}
 (\frac{\dot{R}}{R})^2=\frac{K}{3} (\frac{C_1l}{(l+R)^5} )+\frac{K}{3} (\frac{C_1R}{(l+R)^5})=
\frac{K}{3} (\frac{C_1}{(l+R)^4}).
\end{equation}
and
\begin{equation} \label{10}
\frac{2\ddot{R}}{R}+(\frac{\dot{R}}{R})^2=K((\frac{C_1R}{(l+R)^5} )-p)
\end{equation}
or
\begin{equation} \label{11}
 (\frac{\dot{R}}{R} )^2=\frac{K}{3}(\frac{C_2}{(l+R)^3})
\end{equation}
and
\begin{equation} \label{12}
\frac{2\ddot{R}}{R}+(\frac{\dot{R}}{R})^2=K(\frac{C_2R}{(l+R)^4}),
\end{equation}
respectively.
equation (9) can be rewritten in the form
\begin{equation} \label{13}
\dot R=\sqrt{\frac{KC_1}{3}} (\frac{R}{(l+R)^2}).
\end{equation}
If $l\gg R$, then the right-hand  side of equation(\ref{13}) reduces to the form
$$(\frac{\dot{R}}{R})^2\approx (\frac{K}{3})(\frac{C_1}{l^4}).$$
Its solution represents an exponential function
$$R(t)\approx \exp{\frac{\sqrt{C_1k}t}{\sqrt{3}l^2}},$$
i.e. in the initial period $R$ grows exponentially resembling  the
 extremely rapid expansion of the universe during its inflationary period.
On the other side, if $l\ll R$ then we have
$$(\frac{\dot{R}}{R} )^2\approx \frac{K}{3}\frac{C_1}{R^4}.$$
The solution of this equation represents the time dependence of scale factor, well-known in the standard cosmology
$$R(t)=\sqrt{\frac{2KC_1t}{3}}.$$

Dividing equation(\ref{1}) by square  of the present Hubble constant $H_0^2$ we have
\begin{equation} \label{14}
1=\Omega_0+\Omega_{\Lambda}= \Omega_0(1+\frac{l}{R_0})=\Omega_t.
\end{equation}
Here $$H_0^2=(\frac{\dot R_0}{R_0})^2
, \qquad R_0=1,\qquad \Omega_0=\frac{\rho_0}{\rho_c},\qquad \Omega_{\Lambda}=\frac{L_c}{3H_0^2}=\Omega_0( \frac{l}{R_0}) \qquad
\rho_c=\frac{3H_0^2}{8\pi G}.$$
$\rho_c$ is the critical mass density.

Subtracting equation (\ref{2}) from equation(\ref{1}) we get
\begin{equation} \label{15}
\frac{\ddot R}{R}=-\frac{8\pi}{3}G\rho  +\frac{2}{3}L_c= -\frac{K\rho}{3}(1-\frac{2l}{R}).
\end{equation}
By means of equation (\ref{12}) and equation(\ref{13})  we obtain the deceleration
parameter $q_0$ as a function of $\Omega_0$ and $\Omega_{\Lambda}$ in the form \cite{D}
\begin{equation} \label{16}
 q_0\equiv - \frac{\ddot R R^2}{\dot R^2 R} = \frac{\Omega_0}{2}-\Omega_{\Lambda} =\Omega_0(\frac{1}{2}-\frac{l}{R_0}).
\end{equation}
In our model, the expression of the cosmic age $t_0$ in terms of $\Omega$ and $\Omega_{\Lambda}$ is determined by the integral
\begin{equation} \label{17}
t_0=H_0^{-1}\int\limits_0^1\frac{dR}{R}[\frac{\Omega_0}{R^3}+\frac{\Omega_0l}{R_0}]^{-1/2}.
\end{equation}
The result of the integration of the expression for $t_0$, as a function of $\Omega_0$ and $R_0$,
has the form
\begin{equation} \label{18}
t_0=H_0^{-1}\ \frac{2\arcsin[\frac{l}{R_0}]}{3\sqrt{\Omega_0 l/R_0}}.
\end{equation}
Given $\Omega_0$, $t_0$ becomes only a function of $l$.
Our time-depending $L_c$ predicts  creation of matter at present with an rate of creation per unit volume given by $(C_2\approx M_t)$
\begin{equation} \label{19}
RC=\frac{1}{R^3}\frac{d(\rho_mR^3)}{dt}\mid_0=\frac{4l C_2R_0H_0}{(l+R_0)^5}=\frac{4l\rho_m H_0}{l+R_0}.
\end{equation}

\section{Consequences}
(i) The key role in our further consideration plays the value of $l$.  There are basically two important characteristic lengths
in cosmology which  can be substituted for $l$, (i) Planck's length $l_P=\sqrt{\frac{hG}{c^3}}\approx(10^{-37} cm)$ and (ii) the length assigned to the total mass of the universe
$l_c=\frac{GM_t}{c^2}\approx 2R_0,$
where $R_0$ is the present radius of the universe  $R_0\approx 10^{28} cm$. Taking $l=2R_0$ and $\Omega_0=0.3$ we get,
when inserting this $l$ into equation(\ref{14}), equation(\ref{16}), equation(\ref{18}) and equation(\ref{19}) the following values for the corresponding quantities $\Omega_T\approx 1$,
$t_0\approx 0.98\times H_0^{-1}$,
$q_0\approx -0.5$ and $CR=10^{-47} g^{1} cm ^{-3} s^{-1}$. The later value is about seven order lower than the creation rate in the steady state theory.
 These values are compatible with the estimated observatory ones.\\
(ii) As shown in Fig.(1), if $R=0$ then $\rho_r$ and $\rho_m$  become zero. If $R\gg l$ then $\rho_m(R)\propto C_1R^{-3}$  and $\rho_r(R)\propto C_2R^{-4}$,respectively. The maximum value for $\rho_m$ and $\rho_m$ lies in $l/4$ and $l/3$, respectively.\\
(iii) The acceleration of cosmic evolution continues until the right-hand side of equation(\ref{15}) remains positive then
begins its deceleration.\\
(iv) The fate of the universe in our model is different than the fate nowadays believed.
In a flat or open universe without dark energy, the cosmic expansion continues forever, and
the horizon grows more rapidly than the scale factor.
In our model vacuum energy represents a dynamical variable
which initially drives the expansion of the universe,
later evolves to the present-day small value and in future it becomes continuously smaller converting to zero.\\
(v) Since the mass density assigned to $L_c\propto l_c\rho R^{-1}$ is a dynamical quantity we will determine its mean value over the
whole volume of the universe.
For the total amount of the vacuum energy in the universe, we have
\begin{equation} \label{20}
M_c= \int\limits_0^{R_0}
(\frac{l_c\rho}{R}) 4\pi R^2 dR \approx l_c\rho R_0^2
\end{equation}
Hence, its mean value is $\rho'_c\approx M_c/R_0^3=l_c\rho R_0^{-1}$. With $l_c\approx 2R_0$ we have $\rho'_c\approx 2\rho$.
Recent astronomical observation indicates that the density of the dark energy is small  positive and approximately equal to mass density
 in the universe.\\
(iv) If one takes $l=l_{P}=\frac{Gm_P}{c^2}$ ,where $m_P$ is Planck mass $m_P=\sqrt{\frac{hc}{G}}$ then one has again $\rho_P(t=0)=0$. However, in this case
$R$ grows extremely rapid just on very beginning of cosmic evolution resembling the cosmic inflation. Later, it quickly
starts to evolve very closely to the standard model, so that it has no influence on the present-day universe. The above-considered universe is two-component consisting of the ordinary matter $\rho$ and the energy assigned to $L_c$. To have the accelerating universe, when
taking $l=l_P$, we need an additional component causing  its present acceleration. The simplest way is to add a constant cosmological term  $\Lambda$ to $L_c$. The addition of $\Lambda$ to $L_c$ {\rm does} not change the solutions of equation(\ref{4}) and equation(\ref{7}). It changes the
equations (\ref{1}) and (\ref{2}) in that  $\Lambda$ is added to their right-hand sites . Such three-component universe is accelerating, without initial singularity and with an inflation phase on its beginning.

Finally, we will point to the interesting fact
 that the formula for the total amount of the vacuum energy in the universe given by equation(20) and that calculated  classically within
 the gravito-dynamical field theory (analog to electrodynamics) \cite{R} are similar.
In the gravito-dynamic theory, the analog
to electric charge or charge density  is the expression $\sqrt{G}m$ or $\sqrt{G}\rho$ \cite{IS}, respectively.
Hence, the intensity of gravitational
field  in a sphere with spherically uniform mass distribution whose radius is $R$  is given by the well-known equation
$$\bigtriangledown I_g=-4\pi \sqrt{G}\rho,$$
where $\rho$ is the mass density in the sphere.
Its solution  has the form
$$I_g=-\frac{4\pi }{3}\sqrt{G}\rho R$$
Likewise, as in the case of the electromagnetic field, we determine the field energy density of gravitational field  $E_g$ as
$$E_g=\frac{1}{8\pi} (I_g)^2=\kappa G \rho^2R^2 \qquad \kappa=(\frac{4\pi}{3})^2(\frac{1}{8\pi})\approx 1.$$
If we model the universe  as a sphere filled by the uniform mass density $\rho$ having radius $R_0$ we get for the content
of the field energy-mass
\begin{equation} \label{21}
M_f=\int\limits_0^{R_0}{(\frac{G\rho^2 R^2}{c^2})4\pi R^2 dR}\approx G\rho^2 c^{-2}R_0^5\approx \frac{GM}{c^2}\rho R_0^2=l_c\rho R_0^2.
\end{equation}
We see that the formulas for $M_c$  and $M_f$ are similar. This suggests that the dark energy might represent the field
energy whose source is the mass content of the universe. The field energy can be included into the Einstein equations
through the term $L_c$.

The work was supported by the Slovak Academy Grant Agency VEGA-2/0059/12.


\begin{thebibliography}{99}

\bibitem{TO}
S.  Perlmutter, et al., Astrophys. J {\bf 517}, 565 (1999) 565;
\bibitem{5}
P.J.E. Peebles, in Foundation and Structure in the Universe, edited
by A. Dekel and J.P. Ostriker, Cambridge (1999).
\bibitem{6}
 Y. Yoshii, T. Tsujimoto and K. Kawara, ApJ {\bf 507} (1998), L133.
\bibitem{7}
 G. Steigman and J. E. Felten, Space Sci. Rev. {\bf 74}
(1995),245; G. Steigman, N. Hata and J. E. Felten, ApJ
{\bf 510} (1999), 564.
\bibitem{10} M. Bordag, U. Mohideen and V. M. Mostepanenko, Phys.
Rep. {\bf 353} (2001), 1.
\bibitem{SF}
R.N.Tiwari, S. Ray, Ind. J. Pure. Appl. Math. {\bf 27} (1996) 207;
I. Dymnikova, Class. Quant. Grav. {\bf 19} (2002) 725.
\bibitem{T}
17  P.J.E.Peebles, B. Ratra, Astrophys. J {\bf 325} (1988), L17.
\bibitem{M}
A.S. Al-Rawaf, M.O. Taha, Gen. Relativ. Gravit. {\b 28} (1996), 8; A.S.Al-Rawaf, M.O. Taha,  Phys. Lett. {\bf B 366} (1996), 69;
A.M.M. Abdel-Rahman,  Gen. Relat. Gravit. {\bf 29} (1997), 10;  V. Majernik,  Phys. Lett. {\bf A 282} (2001), 362;
V. Majernik,  Gen. Relativ. Gravit. {\bf 35}, (2003), 6.
\bibitem{WEI}
S. Weinberg, Rev. Mod. Phys. {\bf 61} (1989), 1
\bibitem{4}
L. Knox and L.Page, Phys. Rev. Letts.{\bf 85} (2000),1366.
\bibitem{10} M. Bordag, U. Mohideen and V. M. Mostepanenko, Phys.
Rep. {\bf 353} (2001), 1.
\bibitem{17}
W. Chen and Y-S. Wu, Phys. Rev. {\bf D41} (1990), 695.
\bibitem{MM}
J.A.S.Lima, Alternative Dark Energy Models: An Overview. arXiv: astro-phy/0402109.
\bibitem{D}
B. Ryden, {\it Introdution to Cosmology}. The Ohio State University Press, 2007.
\bibitem{IS}
D. Iwanenko, A. Sokolow, {\it Klassische Feldtheorie.} Akademie-Verlag, Berlin, 1053.
\bibitem{R}
D. Rabounski,  Progress in Physics {\bf 2} (2005), 15.
\end{thebibliography}
\end{document}